# Ontology-Based Users & Requests Clustering in Customer Service Management System[1]

Alexander Smirnov, Mikhail Pashkin, Nikolai Chilov, Tatiana Levashova, Andrew Krizhanovsky, and Alexey Kashevnik

St.Petersburg Institute for Informatics and Automation of the Russian Academy of Sciences, 39, 14th Line, St Petersburg, 199178, Russia
`{Smir, Michael, Nick, Oleg, Aka, Alexey}@iias.spb.su`

**Abstract.** Customer Service Management is one of major business activities to better serve company customers through the introduction of reliable processes and procedures. Today this kind of activities is implemented through e-services to directly involve customers into business processes. Traditionally Customer Service Management involves application of data mining techniques to discover usage patterns from the company knowledge memory. Hence grouping of customers/requests to clusters is one of major technique to improve the level of company customization. The goal of this paper is to present an efficient for implementation approach for clustering users and their requests. The approach uses ontology as knowledge representation model to improve the semantic interoperability between units of the company and customers. Some fragments of the approach tested in an industrial company are also presented in the paper.

**Keywords:** Text Mining, Clustering, Ontology, Agent

## 1 Introduction

Many research efforts have been undertaken in the area of Customer Service Management (CSM) to perform a shift from "product-centric" production to "customer-centric" production. CSM has the following main functions: searching for information about company's products by the customers (with clustering, ranking, etc. to organize found results); storing, organizing and processing information about the users and their contacts by the system administrators and managers [1].

Developed by the authors CSM system is referred to as "Intelligent Access to Catalogue and Documents" (IACD) and has the following major features: (i) intelligence of the system in providing interface forms: static templates for special

---

structured inputs and precise results for specific tasks, free text inputs for knowledge sources search, and learning-based intelligent adviser; and (ii) customizability: from unknown unspecified customer to building and supporting target groups (e.g., by job titles, area of interests etc.), and to personalized profile-based support. Structured customer requests represent templates (specially designed forms for searching within a limited group of products/solutions) that allow achieving high relevance of the found results but miss universality. Free text requests have maximal universality but achieving high levels of the result relevance is a challenging task. Described here CSM system does this by setting some syntactical constraints on the free text requests and by using a part of the shared ontology of the company. To further improve free text request processing it is reasonable to accumulate information about customers' interests by grouping based on user profiles (with request history information) and using text mining techniques. These are the topics the paper concentrates on.

Grouping can show similarities between different customers that would make it possible to better serve them, to provide interesting for them information "just-in-time" or even "just-before-time". Besides, producing "good" groups can provide additional useful benefits (e.g., better filtering of results corresponding to customers' interests). For this purpose clustering of the customers into a number of distinct segments or groups in an effective and efficient manner is required. Text clustering (direction of text mining [2]) helps in customer problems (interests or preferences) identification and classification.

There are the following ontology-based clustering scenarios suitable for CRM:

First, *requests clustering for one user*. The goal of this clustering is to define prevalent user interests (e.g. product category, preferred brand name, level of quality). This is important to foresee user's needs and satisfy it just in time.

Second, *users clustering*. This scenario identifies users groups. Without information about customer groups the CSM system administrator should treat each user separately. On the contrary, treating only of each group (after the clustering) will be less time-consuming, because number of groups is much smaller than number of users.

Third, *requests clustering*. This scenario groups together similar requests. This clustering is used in order identify categories of users' interests. Another goal of requests' clustering is to perform text analysis to get 1) common types of user request, 2) common misspellings, 3) customer/user lexicons, 4) frequency / popularity of used term, 5) bottleneck requests (requests that return too many / few results), 6) promising request (i.e. potential customers/new market/direction of future work), 7) recommendations for request templates development, etc.

Agents are very promising technology for distributed data / text mining [3], [4].

The paper is structured as follows: section 2 describes research efforts related to the system IACD and implemented scenarios. In section 3 a description of the clustering algorithm is presented. Experiments and future work discussions conclude the paper.

## 2 The System IACD: The Concept and Functions

### 2.1. KSNet-Based Customer Service Management

Recently, there has been an increased interest in developing CSM systems that incorporate knowledge management and data & text mining techniques [1].

Proposed by the authors KSNet-approach considers knowledge logistics (a direction of knowledge management) as a problem of a knowledge source network (KSNet) configuration that includes as network units - end-users / customers, loosely coupled knowledge sources / resources, and a set of tools and methods for information / knowledge processing [5]. A multi-agent system architecture based on FIPA Reference Model was chosen as a technological basis for the KSNet-approach. FIPA-based technological kernel agents used in the system are: wrapper (interaction with knowledge sources), facilitator ("yellow pages" directory service for the agents), mediator (task execution control), and user agent (interaction with users).

The KSNet approach is selected as a kernel for creation a distributed CSM system which provides the global company's face to the customer through a single point of information access [6]. Major motivation of this solution was that the agent-based technology is a good basis for CSM in the global companies since agents can operate in a distributed environment independently from the user and apply ontologies to knowledge representation, sharing and exchange.

Ontology could be specified as a set of concepts with informal definitions, a set of relations holding among these concepts not limited to hierarchical ones (*is-a* and *part-of*), and a set of axioms to formalize the definitions and relations [7]. Here ontology plays very important role as a common vocabulary (language) in agent community and the company, and as a model which supports semantic interoperability of company units (customers, users, departments, plants, etc.). User profiles are used during interactions to provide for an efficient personalized service.

The system is considered as an Internet-based support system where customers initiate a real-time "electronic dialog" with the customer support agent. For each dialog session, the dialog between agent and customer interaction is recorded and stored in the customer / user profile. The structured elements of this session log include information about the customer (who, what, where, when, etc.). The unstructured data is the verbatim (free-form text) of the customer/agent dialog itself. Timely and accurate customer & request group identification is critical for the support agents as well as the product engineers and service managers. The system is an example of deriving value from the integration of free-form text (dialogs and user requests) and structured data (electronic product catalogues, etc.).

The aim of text mining is similar to data mining in that it attempts to analyze texts to discover interesting patterns such as clusters, associations, deviations, similarities, and differences in sets of text [8]. Text mining process consists of six steps — source selection, information retrieval from text collection, information extraction to obtain data from individual texts, data warehousing for the extracted data, data mining to discover useful pattern in the data, and visualization of the resulting pattern [9]. Here major techniques include clustering and classification methods, such as nearest neighbor, relational learning models, and genetic algorithms, and dependency models, including graph-theoretic link analysis, linear regression, decision trees, nonlinear regression, and neural networks [10].

Ontology-based CSM systems could help to determine characteristics of the customer / user data and of the desired mining results, and to enumerate the knowledge discovery processes that are valid for producing the desired results from the given company data / text sets. The ontology-based approach was developed for CSM and implemented in the system IACD. Current version of the system provides for customers a common way to search for products and solutions and presents information about different applications: (i) technical data of company's products, (ii) project-specific solutions based on tasks' conditions given by customers and (iii) corporate documents and available Web sites taking into account customers' interests and constraints stored in the corporate ontology. It helps to easily find solutions for planning simple methods and for alternative comparison. Especially ontology implementation is useful when prediction of customers' interests is required but customers use different languages, different terms, different levels of abstraction and different units of measures, have different areas of interests and different levels of customizability, and make decisions on different levels.

User profiles are used heavily in clustering (customer & request grouping). For better customer serving, the approach assumes creation of user profiles correlating with the ontology. User profile, besides other information and knowledge characterizing the user and the user activity, stores the history of the user requests. The user requests are grouping on the basis of the similarity metrics and hierarchical relations of ontology classes.

Referring to text mining steps some related to CSM activities problems could find solution based on the ontology-driven approach. The first problem is that the results may differ from the real user needs. The second problem is that the description of the sources is not uniform – there are different formats such as sentences, items and tables.

It is difficult to extract the information by conventional methods of information extraction [11]. Filtering agent / service based on domain expert knowledge is needed (ontology could be useful here too). Current methods in text mining use keywords, term weighting, or association rules to present text context. It can be implemented by means of *context filtering*.

The following classical text mining techniques are also used in the system IACD: text processing and clustering [Fig. 1]. *Text processing* includes a sequence of steps: tokenization, stop-words finding, spelling, stemming, search for units of measures (e.g. "kg", "mm"), etc. [12]. The following problem-oriented agents specific for KSNet, and scenarios for their collaboration were developed and adapted for text mining problems: (i) text processing — translation agent (terms translation between different vocabularies) and ontology management agent (ontology operations performance) and (ii) clustering — monitoring agent (verification of knowledge sources).

Input for clustering algorithm would be user requests and the ontology. In [2] the following categorization of major clustering methods is proposed: partitioning methods, hierarchical methods, density-based methods, grid-based methods, model-based methods. The hierarchical method is used in the proposed approach that is described in detail in sec. 3. Since user requests and instances of ontology classes can be tied with context information (e.g., time, location, language), input data could be filtered in order to mine data about products which are geographically located near customer, delivered in time, have appropriate level of quality.

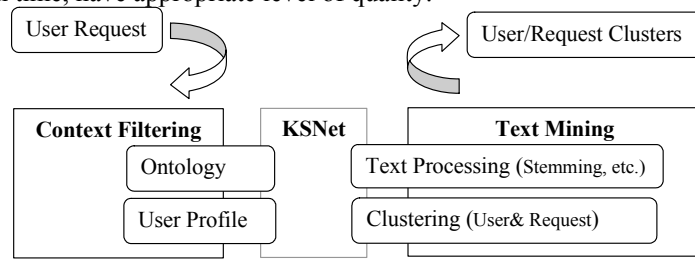

**Fig. 1.** Application of the KSNet-approach to ontology-based CSM

### 2.2 The System IACD Functional Specification

The system IACD has been developed for a company producing manufacturing equipment that has more than 300.000 customers in 176 countries supported by more than 50 companies worldwide with more than 250 branch offices and authorized agencies in further 36 countries.

The main goal of the presented here CSM system based on the KSNet-approach is to provide information about solutions to customers in addition to existing product catalogue and to find products and solutions. Therefore, besides company's documents two other applications were selected as knowledge sources: (i) the product catalogue containing information about 20'000 items produced by the company: technical data, price, etc. for different languages, and (ii) an application containing a set of rules for configuration of handling system projects, structured data for industry segment and automation function description, and technical data of carried out products. These $1^{st}$ and $2^{nd}$ applications are oriented to industrial engineers and designers. Extension of these target groups with new ones allows increasing the number of potential clients and providing additional benefits to the company. Based on this information a part of the shared ontology of the company is built. The ontology uses frame-based knowledge representation model and includes classes and attributes.

Usually, it is proposed to have a shared ontology for a global company. However, the practice shows that this is not always possible due to the large number and heterogeneity of company members. Sometimes it is enough to build a smaller shared ontology for one aspect of the company activities or one company member only (later in the paper referred to as "ontology"), but this ontology should also support synonyms that might be used by its customers to provide for interoperability. This is how it was implemented in the presented approach.

The developed CRM system uses the company ontology consisting of more than 240 classes, 355 attributes from different sources [12]. The ontology currently is based on six taxonomies: VDMA (association of German machine and plant construction companies), industry segments (with basic functions as attributes – Work-piece, Handling, Assembly, Light assembly, Packaging, Automotive processes, Process automation, Food manufacturing, etc.), automation functions (with basic functions as attributes), handling system projects classification, technical data of industrial applications, and user-defined taxonomy.

The following scenario of the customer access to corporate information was developed: the customer passes authentication procedure, selects an appropriate interface form and enters a request into the system. The system recognizes the request and defines which data the customer needs. If the customer needs to solve a problem the system defines load conditions (parameters describing a certain problem: e.g. mass to be moved, direction of the transportation, environmental conditions etc.) and looks for handling system projects. If no certain problem is defined by the customer the system checks information in product catalogue, database storing technical data of standard handling systems and in company's documents.

*Implemented approach for the user and request clustering scenarios* (see 1st, 2nd and 3rd scenarios in the introduction) has the following steps:

1. Extract words / phrases from the request;
2. Calculate *similarity metrics* between the request and ontology (i.e. compare text strings extracted from the request and the name of the class or attribute);
3. Ontology-based algorithm of users & requests clustering
   Step 1: Construct weighted graph consisting of nodes: classes, attributes, and users. Weights of arcs are calculated on the basis of 1) similarity metrics (i.e. they are different for different user requests) and 2) taxonomic relations in ontology;
   Step 2: Construct weighted graph consisting of user / customers (when classes and attributes are removed, arcs weights are recalculated).
   Step 3: Hierarchical clustering of users (customers) graph.

## 3 Ontology-Based Algorithm of Users & Requests Clustering

### 3.1. Similarity Metrics

Calculation of similarity between the user request and the corporate ontology is based on the similarity of the names for classes and attributes of the ontology and strings (concepts, phrases) of the request.

*Request-Class Similarity.* The essence of this task is to find classes in the ontology corresponding to the user request. This is done by comparing names of the classes (text strings) with stemmed and corrected (when misspelled) words extracted from the user request (names of classes, which are not numbers, misspelled words or units of measures).

The algorithm of *fuzzy string comparison* is used for this purpose. It calculates occurrence of substrings of one string in the other string. The algorithm can be illustrated by comparing strings "motor" and "mortar".

The first string "motor" has 5 different substrings (m, o, t, r, mo) contained in the second string "mortar". The total number of different substrings in "motor" is the following 13 strings: ((m, o, t, r), (mo, ot, to, or), (mot, oto, tor), (moto, otor)). The result is the following string "motor" corresponds to the string "mortar" with the similarity of 5/13 or 38%.

*Request-Attribute Similarity.* Attributes (corresponding to the user request) are searched within the names of the ontology elements. Regular expressions [13] are used for text processing. Below, *Entries* denotes a part of attribute name (one or several words) found in the request:

*Request*: "Pay load 5 kg, *Stroke X* 100 mm, *Stroke* Y 200 mm"
*Attribute*: "Stroke X"
*Entries (parts of user requests)*: "Stroke X", "Stroke"

In this example the attribute name "Stroke X" (from the ontology) is found in the request once entirely ("Stroke X") and once partially ("Stroke").

For each attribute the following parameters are defined: $N_{Words}$ — number of words in the attribute name; $AttrName_{Rest}[0..N_{Words}]$ — array of words forming the attribute name; $Word_{Attr}$ — an element of the array $AttrName_{Rest}[]$; $Position_{UserRequest}$ — point to the current position in the user request.

The algorithm of calculating *the similarity of the attribute to the request* is the following:

```
N_Entries = 0;
Position_UserRequest = 0;
Initialize AttrName_Rest[];
Similarity = 0;
FOREACH Word_Attr IN (AttrName_Rest[]) {
  FOREACH (Position_New =
  strpos(Word_Attr, UserRequest, Position_UserRequest))
  {
    Position_UserRequest += Position_New;
    remove Word_Attr from AttrName_Rest[];
    Similarity = max (
        Similarity,
        CalcSimilarity(Position_UserRequest, AttrName_Rest[]));
  }
}
```

Similarity for each part of the request (entry) to the attribute is calculated by the function *CalcSimilarity()* with the following properties. First, the more words from a name of the attribute are found in the request, the greater similarity. For example, *CalcSimilarity()* is greater for the entry "Stroke X" than for the entry "Stroke". Second, the longer the *Entry* the greater the similarity, e.g. similarity of "Stroke" is greater than that of "X".

Function *strpos* in the algorithm above returns the position of the attribute substring $Word_{Attr}$ in the string UserRequest, starting at the position "$Position_{UserRequest}$". So, the variable $Position_{UserRequest}$ takes values (starting from zero) through all the positions of the *attribute* substring in the text of the *user request*.

Thus, the similarity metric shows the degree of the correspondence of user requests to classes and attributes of the ontology. Similarity is a real number in the range [0, 1]. After processing the user requests, extracting terms, calculating similarity metrics, an XML structure is filled. It consists of the tags (i) related to classes

(<CID> — the ontology class ID (unique identifier), <CWeight> — similarity of the class to the user request) and (ii) the tags related to attributes (<AID> — ID of an attribute from the ontology, <AWeight> — similarity of the attribute to the user request). These data are the input for the clustering algorithm described in the next section.

### 3.2. Users & Requests Clustering

This section describes proposed ontology-based clustering algorithm related to agglomerative hierarchical methods. The hierarchical method creates a decomposition of the given set of data objects. The agglomerative approach, also called the *bottom-up* approach, starts with each object forming a separate group. It successively merges the objects or groups close to one another, until all of the groups are merged into one (the topmost level of the hierarchy), or until a termination condition holds. More information about hierarchical methods see in [2].

The proposed algorithm relies on user profiles consisting of: (i) personal data (user name, country, etc.); (ii) list of classes (found in the ontology using requests of the user) and their similarity to the user/request $CU_{sim}$; (iii) list of attributes (found in the ontology using requests of the user) and their similarity to the user/request $AU_{sim}$. It is needed to group the users by their requests. A groups should not consist of all the users or (another extreme case) of only one user.

The user in this task is presented via a set of his/her requests. Therefore it is possible to turn from "request" relations (class-request and attribute-request) to "user" relations (class-user (CU) and attribute-user (AU)).

A weighted user-ontology graph $G_0$=<N, E>=<(C, A, U), (CA, CC, CU, AU)> is considered. N represents three types of nodes: C — class, A — attribute and U — user. Since arcs E=(CA, CC, CU, AU) of the graph $G_0$ are marked with numbers (weights) then graph $G_0$ can be presented as *weight matrix*:

$$G_0[i, j] = \begin{cases} 0, if\ i = j \\ c_{ij}, \text{finite quantity, if there is an arc from node i to j} \\ \infty, \text{if there is no arc from i to j} \end{cases} \quad (1)$$

User-ontology graph $G_0$ consists of two types of arcs. The type I of arcs (CA, CC) is defined by the hierarchy of classes and belonging to them attributes in the ontology. The type II of arcs (CU, AU) is defined by the set of relations between use requests and classes/attributes (Fig. 2).

Weights of arcs between the nodes representing classes and users $CU_{weight}$, and attributes and users $AU_{weight}$ are defined via the similarity $CU_{sim}$ and $AU_{sim}$ (values of XML tags <CWeight> and <AWeight>) as follows:

$$CU_{weight} = 1 - CU_{sim} \quad (2)$$
$$AU_{weight} = 1 - AU_{sim}$$

All arcs CA and CC tying together classes and attributes have $CA_{weight}$, $CC_{weight}$ $\in (\varepsilon, 1)$ defined by the IACD system administrator (see experiment results in sec.

4). $CC_{weight}$ denotes arcs' weight of linked classes in the ontology. $CA_{weight}$ — arcs' weight of linked attributes and classes.

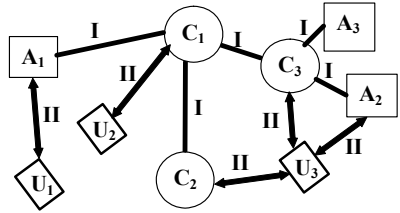

**Fig. 2.** User-ontology graph $G_0$

*Step 1: Weight Assignment.* The weight of the arc CU (class–user/request) is considered in the graph $G_0$. The class C can be encountered several times in different requests of user U in this graph. This should be taken into account in the weight of the arc CU. If class C is encountered in N requests of user U, then the *weight of arc* $CU_{weight}$ between class C and user U can be defined as follows:

$$CU_{weight} = 1 - \sum_{i=1}^{N} \frac{1 - CU_{weight}^{i}}{N_{max}}, \quad (3)$$

where $N_{max}$ is the maximum number of instances of the same class for all requests.

If similarity $CU_{sim}$ is equal to 1 then the weight of arcs CU (AU) will be equal to 0 (see (2)) and the user with several similar requests will not have priority. So, in order to take into account the number of requests which coincide with the same class (attribute) several times, weight $CU_{weight}$ and $AU_{weight}$ should be assigned to $\varepsilon$ instead of 0. $\varepsilon$ should be close enough to 0 and defined in the range (0,1).

*Step 2: User Graph Constructing.* Usually Floyd algorithm [14] is used to find the shortest path between every pair of the graph nodes. In the considered case it is enough to know the weight of the *shortest path* between every pair of users. Therefore the *modified Floyd algorithm* is proposed and used. The table T of relations between users and classes is defined by the user-ontology graph:
- T[i, j] is the weight between nodes i and j,
- T[i, j] = ∞, if there is no path from node i to j.

The variable p in the algorithm is the number of nodes in the graph $G_1$.

Algorithm to calculate the weight of the shortest path between every pair of users in the graph $G_1$:

```
for i from 1 to p do
 for j from 1 to p do
  for k from 1 to p do
   if i<>j and T[i,j]<>∞ and
      i<>k and T[i,k]<>∞ and
            (T[j,k]= ∞ or
             T[j,k]>T[j,i]+T[i,k])
```

```
                then
                        T[j,k] := T[j,i]+T[i,k]
            end if
        end for
    end for
end for
```

As a result the matrix T is created (i.e. users' graph $G_2$). Thus the weight of arc between every pair of users in $G_2$ is equal to the weight of the shortest path in $G_1$, or ∞ if such path does not exist.

*Step 3: User Graph Clustering.* Now in order to group users to clusters it is enough to divide the graph $G_2$ to subgraphs $G^i$, i=1,n, where n – is the number of clusters. Cluster mass $G^i$ is the sum of weights of all arcs in the subgraph $G^i$.

The optimal clustering is proposed to be defined as follows:
a) n→min, i.e. minimize number of user groups,
b) $D_{max} > D[G^i]$, i=1,n, i.e. the maximum cluster mass for every subgraph is less than some (defined in advance by the IACD system administrator) constant $D_{max}$.

Hierarchical clustering algorithm:

```
1. D[U_i]=0, i=1,n. At the beginning of the algorithm
   every node corresponds to a subgraph. The value of
   subgraph mass D[U_i] is zero.
2. Fill vector A: A[z] = ARC_weight + D[U_i] + D[U_j]; i.e.
   element A[z] of vector A equals to the sum of weight
   of the arc between U_i and U_j nodes of the graph and
   cluster mass D[U_i] + D[U_j] of these nodes.
3. Take the element A[z] from the vector A with the mini-
   mum value (sum of weight of the arc and the cluster
   mass).
4. If A[z] > D_Max then terminate algorithm.
5. Join nodes U_i, U_j; result mass of U_i, U_j is D[U_i] =
   ARC_weight[i,j] + D[U_i] + D[U_j]; remove D[U_j] from vector
   D and the minimal element from A.
6. Update values in the vector A for arcs of adjacent
   nodes U_i (e.g. if node U_k is an adjacent node to U_i
   then A[ik] = ARC_weight[i,k] + D[U_i] + D[U_k])
7. Go to line 3.
```

Due to the testing of inequality $D < D_{max}$ (line number 4), the second condition of the optimal clustering (b) will be satisfied.

Algorithm time complexity uses three parameters (N – sum of classes and users (class-user graph size); n — number of users (user graph); L – number of classes related to a request) and has three constituents. First, time complexity of creating class-user graph is $o(N \cdot L)$, because every L classes/attributes are retrieved from a database for each N requests. Second, modified Floyd algorithm uses three nested loops to calculate the shortest paths between nodes in graph, so its complexity is $o(N^3)$.

Third, time complexity of hierarchical clustering algorithm has two parts: outside of the cycle (steps 1-2) and of the cycle itself (steps 3-7). Complexity of step 1 is $o(1)$ and step 2 is $o(n^2)$ in case of building vector $A$ for fully connected graph. The cycle: step 3 (searching minimal element in vector $A$) is $o(n)$, step 4 and step 5 has complexity $o(1)$, step 6 is $o(n)$ for fully connected graph (vector $A$ is updating for adjacent arcs). Number of maximum cycle iterations is n in the worse case (one cluster includes all users). So, time complexity of the cycle is $o(n^2)$. Thus, complexity of hierarchical clustering algorithm is $o(n^2) + o(n^2) = o(n^2)$.

So, total time complexity of the three constituents is $o(N \cdot L) + o(N^3) + o(n^2)$. It is $o(N \cdot L) + o(N^3)$ when N ≥ n. Usually number of requests N is greater then number of classes L related to a request due to user requests are short enough (L ≤ N), so $o(N^2) + o(N^3) = o(N^3)$.

## 4  Experiments

The goal of these experiments is to study the clustering approach and to evaluate how user clusterization depends on variables $D_{Max}$ and $CC_{weight}$. The IACD system administrator can use this dependence to determine parameters of the clusterization ($D_{Max}$ and $CC_{weight}$).

The graph (Fig. 3) demonstrates automatically generated set of two request clusters using the Graphviz tool [15]. The following parameters are used for this clustering: number of classes = 10, number of users = 5, $CC_{Weight}$ = 0.2 for all the classes in the ontology and $D_{Max}$ (maximal cluster mass) = 0.6. For illustrative purpose user requests are shown instead of users. Classes in Fig. 3 are marked with ovals and requests are marked with rectangles. Clusters are marked with rectangles enclosing requests' rectangles. The magnified part of the graph includes: classes: "Pick & place" and "Projects", and requests: "Pick & place" and "Pick & place pay load > 0 and Stroke = 5". The exact name of the class "Pick & Place" is presented in both requests, so arc weight between the request and the class "Pick & Place" is $\varepsilon$ =0,001 (Fig. 3) in both cases (two arcs from class Pick & Place to these requests). $CC_{Weight}$ = 0.2, hence the weight of arc between the class "Pick & place" and the class "Projects" is 0.2.

It can be seen (Fig. 4) that for $D_{Max} < \varepsilon$ = 0,0001 the number of groups is equal to the number of users for all values $CC_{Weight}$ (AB in Fig. 4). Increasing $D_{Max}$ causes the number of groups to decrease down to two ($BC_i$ in Fig. 4). Then, the steady level of the group number ($C_1D_1$ and $C_2D_2$ in Fig. 4) can be seen. At this interval the groups remain unchanged though $D_{Max}$ increases. Given $CC_{Weight} = \varepsilon$ the interval $C_3D_3$ is lacking. This can be explained as follows. If the value of $CC_{Weight}$ is low, to add a new user to the cluster, $D_{Max}$ has to increase by the value equal to the weight between that user and the nearest to him/her class. In this case weight between $C_2$ and $C_5$ can be neglected because it is significantly lower than weight between that user and the class nearest to him/her. Area $D_iE$ describes a low number of groups (tending to one group consisting of all users). It can be concluded that the IACD system administrator has to define $D_{Max}$ from within $C_iD_i$ interval.

**Fig. 3.** The user graph consists of two clusters which are marked by rectangles. A part of this graph is zoomed in for convenient reading

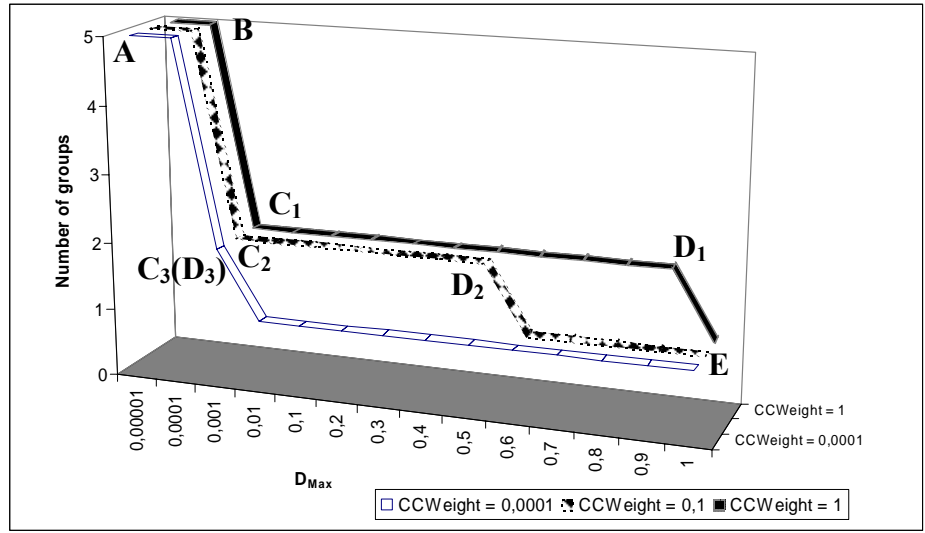

**Fig. 4.** Number of clusters depending on $D_{Max}$ and $CC_{weight}$

## 5   Discussion and Conclusion

Ontology-based clustering is used to group related users and request information. Difference of this approach and other approaches are discussed below.

In [16] the ROCK hierarchical clustering algorithm for categorical attributes is proposed and implemented. In ROCK concept of links to measure the similarity/proximity between a pair of data points is presented. The feature of the ROCK algorithm is that the merging clusters/points with the most number of links is performed in the first order during the clustering. In contrast, the clustering algorithm proposed in this article merges clusters/points which are nearest in the graph. Another distinctive feature of the algorithm proposed in this article is that text processing and ontology data are used to form input data for clustering.

In [17] COSA (Concept Selection and Aggregation) the approach uses a simple, core, domain-specific ontology for restricting a set of relevant document features to cluster documents by K-Means algorithm. The definition of clustering quality metric is one of the benefits of this work. The user may decide to prefer one over another clustering result based on the actual concepts used for clustering as well as on standard quality measures (such as the silhouette measure). Unlike proposed here approach considering relatively short user requests, in [17] clustered data are text documents. Another difference is that the quality of the result is estimated by the administrator in the proposed approach.

Human-based and computer based clustering methods is one more classification of clustering. For example, in [18] an integration of the technical domain experts work with the data mining tools is presented. In this integration the following step is done to cluster science documents: (1) defining of words frequency (to identify topic of document), (2) determination of the relationships among themes, and (3) tracking the evolution of these themes and their relationships through time. The first and the second step are designed and implemented (see [18]). Since in this paper the definition of parameters of ontology-based clustering algorithm by the system administrator is very crucial for the algorithm (sec. 4), it can be concluded that the approach presented here is integrated (human-based & computer based).

In [19] the web-users clustering algorithm (based on K-means clustering and genetic algorithm) is proposed, implemented, and tested. The web-users clustering algorithm uses the integration of a neural network and self-organized maps (SOM) to cope with insufficient information about users. In [4] ideas about distributed data clustering are presented. This distributed data clustering is density-based and takes into account the issues of privacy and communications costs. Since genetic algorithms, SOM, and density methodic are not used in the proposed approach, these works can not be compared with.

Possible improving of the implemented in the system IACD clustering approach can be achieved by taking into account probability relations between classes and attributes in the ontology. It is an ontology engineering task that requires involvement of domain experts. This type of ontology can be very suitable for semantic relations in text mining area, because complex semantic relations can not be reduced to simple present/absent relations (like in ontology taxonomy). This is believed to increase the efficiency of context-aware CSM system especially for learning and predicting customer actions and interests, better recognition of their requests, etc.

The system IACD has the following properties of modern applications [Chen, 2001] related to free text processing (which is necessary for customer request recognition): (i) *multi-language support*, currently the system IACD supports three languages: English, Russian, German; (ii) *automatic taxonomy creation*, the company ontology (taxonomy) is built automatically in the system IACD based on the available knowledge sources; (iii) *domain-specific knowledge filter* (using vocabularies or ontologies), four level ontology is used in IACD; (iv) *indexer*: all documents are indexed for rapid access; (v) *multi-document format support*, the system IACD supports MSOffice documents, RTF documents, web pages, Adobe PDF files, ASCII text files; (vi) *natural or statistical language processing*; in IACD natural language processing consists of tokenization, spelling, stemming, etc.; (vii) *term extraction*, in IACD names of ontology classes and attributes, units of measures (e.g. "kg" and "mm") are extracted from customer requests.

## Acknowledgment

Some parts of the research were done by the Contract titled "Intelligent Access to Catalogues and Documents" between Festo and SPIIRAS and as parts of the project # 16.2.44 of the research program "Mathematical Modelling and Intelligent Systems" and the project # 1.9 of the research program "Fundamental Basics of Information Technologies and Computer Systems" of the Russian Academy of Sciences.